\documentclass[prl,twocolumn,twoside,preprintnumbers,superscriptaddress,showpacs,showkeys,nofootinbib,groupedaddress]{revtex4-1}

\usepackage{graphicx}

\usepackage{dcolumn}
\usepackage{bm}
\usepackage{color}
\usepackage{amsmath,amssymb}

\begin{document}

\title{
  Double Gamow-Teller transitions
  and its relation to neutrinoless $\beta\beta$ decay
}

\author{Noritaka Shimizu}
\email{shimizu@cns.s.u-tokyo.ac.jp}
\address{
	Center for Nuclear Study, 
	The University of Tokyo, 
	Bunkyo-ku, Hongo, Tokyo 113-0033, Japan 
}

\author{Javier Men\'endez}
\email{menendez@cns.s.u-tokyo.ac.jp}
\address{
	Center for Nuclear Study, 
	The University of Tokyo, 
	Bunkyo-ku, Hongo, Tokyo 113-0033, Japan 
}

\author{Kentaro Yako}
\address{
	Center for Nuclear Study, 
	The University of Tokyo, 
	Bunkyo-ku, Hongo, Tokyo 113-0033, Japan 
}

\date{\today}

\begin{abstract}
    We study the double Gamow-Teller (DGT) strength distribution of $^{48}$Ca 
    with state-of-the-art large-scale nuclear shell model calculations.
    Our analysis shows that the centroid energy of the DGT giant resonance
    depends mostly on the isovector pairing interaction,
    while the resonance width is more sensitive to isoscalar pairing.
    Pairing correlations are also 
    key 
    in neutrinoless $\beta\beta$ ($0\nu\beta\beta$) 
    decay. We find a simple relation between the centroid energy 
    of the $^{48}$Ca DGT giant resonance and 
    the $0\nu\beta\beta$ decay nuclear matrix element.
    More generally, we observe a very good linear correlation
    between the DGT transition to the ground state of the final nucleus
    and the $0\nu\beta\beta$ decay matrix element.
    The correlation,
    which originates on the dominant short-range character of both transitions,
    extends to heavier systems
    including several $\beta\beta$ emitters,
    and also holds in energy-density functional results.
    Our findings suggest that DGT experiments
    can be a very valuable tool 
    to obtain information on the value of
    $0\nu\beta\beta$ decay nuclear matrix elements.
\end{abstract}


\maketitle

{\it Introduction.}
The quest for the detection of
neutrinoless $\beta\beta$ ($0\nu\beta\beta$) decay
is one of major experimental challenges
in particle and nuclear physics~\cite{KamLAND-Zen16,GERDA17,EXO14,CUORE15}.
$0\nu\beta\beta$ decay is the most promising process
to observe lepton-number violation in the laboratory.
Its discovery would prove that neutrinos are its own antiparticles
(Majorana particles),
provide information on the absolute neutrino mass,
and give insight on the matter-antimatter asymmetry
of the universe~\cite{avignone}.
The $0\nu\beta\beta$ decay lifetime depends on the
nuclear matrix element (NME),
which has to be calculated 
and is sensitive to the nuclear structure
of the parent and daughter nuclei.
Nuclear many-body approaches disagree in their prediction
of NMEs by more than a factor two.
Furthermore these results may need an additional
renormalization, or ``quenching'', of a similar amount~\cite{engel-ropp}.
This theoretical uncertainty limits severely the capability
to anticipate the reach of future $0\nu\beta\beta$ decay experiments,
and the extraction of the neutrino mass
once a decay signal has been observed.

Given the difficulty of theoretical calculations
to agree on the value of the $0\nu\beta\beta$ decay NMEs,
experimental data on the nuclear structure
of the parent and daughter nuclei~\cite{freeman-12,kay-13,szwec,entwisle},
two-nucleon transfer reactions~\cite{brown-14},
or the lepton-number-conserving two-neutrino $\beta\beta$ decay~\cite{barabash-15,caurier-90,horoi-16,rodin-06,suhonen-12,barea-15}
have been proposed to test the many-body approaches
and shed light on the NME values.
Charge-exchange reactions,
where a proton is replaced by a neutron or the other way around,
provide information on the Gamow-Teller (GT) strengths~\cite{yako-ca48,freckers-13,ichimura-06},
offering another good test
of the theoretical calculations~\cite{poves-12,suhonen-13,rodriguez-11}.
The GT strength of the parent to intermediate nuclei, combined with the strength of the daughter to intermediate nuclei, is related to
two-neutrino $\beta\beta$ decay.
However the connection is not straightforward 
because the relative phase between the two GT contributions
cannot be measured.
Despite all these efforts 
an observable clearly correlated to
$0\nu\beta\beta$ decay remains to be found.

Double charge-exchange reactions
have been suggested to resemble
$0\nu\beta\beta$ decay~\cite{rodriguez, cappuzzello}.
The detection of the resulting new collective motion is, however, challenging.
While the double isobaric analogue resonance was found
via pion double charge-exchange reactions~\cite{pion-dcx},
the double Gamow-Teller giant resonance (DGT GR)
remains to be observed 
three decades after the first detailed theoretical predictions
~\cite{auerbach-first-dgt,vogel-sumrule,muto-sumrule,zheng,zheng-sumrule}.
A more recent study was carried out in Ref.~\cite{sagawa-sumrule}.
Modern searches of the DGT GR are
based on novel heavy-ion double charge-exchange 
reactions
~\cite{takaki-aris,takahisa-17}.
The data analysis of experiments performed at RNCP Osaka
are recent~\cite{takahisa-17} or ongoing~\cite{takaki-cns},
and a similar experiment
is planned at RIBF RIKEN~\cite{uesaka-nppac}.
Double charge-exchange reactions will be used
at LNS Catania aiming to give insight on
$0\nu\beta\beta$ decay NMEs~\cite{cappuzzello}.

\begin{figure*}[t]
	\centering
	\includegraphics[width=0.88\textwidth]{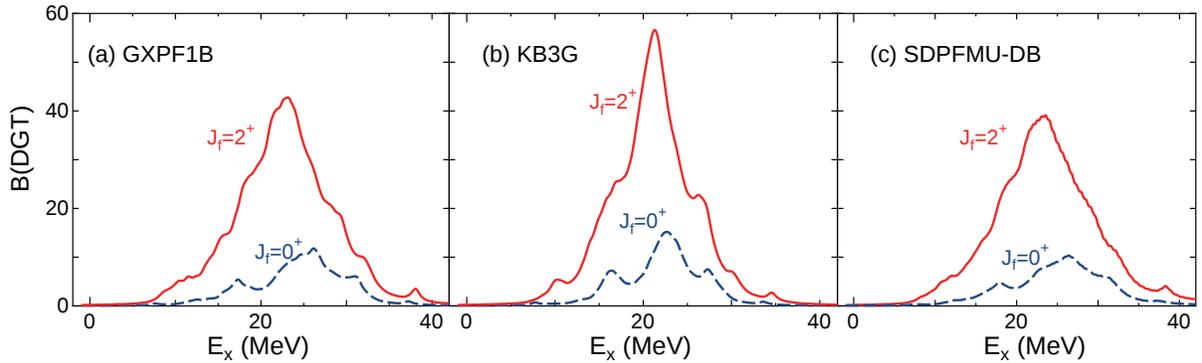}	
	\caption{
          DGT strength distribution of $^{48}$Ca to $^{48}$Ti.
          The red solid  (blue dashed) line shows the
          $B({\rm DGT}^-;\lambda=2)$ [$B({\rm DGT}^-;\lambda=0)$] transitions.
          Results obtained with the (a) GXPF1B \cite{gxpf1b}, (b) KB3G \cite{kb3g} nuclear interactions in the single $pf$ shell, and (c) SDPFMU-DB~\cite{iwata-prl2016} interaction (an extension of GXPF1B)
          in the two-shell $sd$-$pf$ configuration space. 
	}
	\label{fig:dgt-interactions}
\end{figure*}

Present DGT GR searches use
the lightest $\beta\beta$ emitter $^{48}$Ca as target.  
Several experimental~\cite{nemo-3,candles,carvel}
and theoretical works~\cite{iwata-prl2016,song-17,barea-15,vaquero-13,ca48-bb-mass,
	horoi-nonclosure-48ca,menendez-09,simkovic-13}
have investigated
the $0\nu\beta\beta$ decay of this nucleus.
In this Letter we analyze the $^{48}$Ca
DGT strength distribution in connection to $0\nu\beta\beta$ decay.
We focus on the correlation between the properties of the DGT GR
and the $0\nu\beta\beta$ decay NME.
In addition we study the relation between
the DGT transition to the ground state of the final nucleus 
and the NME, 
exploring the extension to heavier $\beta\beta$ decay emitters.

{\it Double Gamow-Teller transitions.}
\label{sec:sumrule}
The DGT transition probability, or strength, is defined as
\begin{equation}
\label{eq:bdgt}
B(DGT^{\pm}; \lambda;i\rightarrow f) = 
\frac{1}{2J_i+1}| \langle f || {\cal O}^{(\lambda)}_{\pm} || i \rangle |^2,
\end{equation}
with $J_i$ the total angular momentum of the initial ($i$) nucleus.
The DGT operator couples two GT ones
${\cal O}^{(\lambda)}_{\pm} =[ \sum_j \bm{\sigma}_j \tau^\pm_j 
\times \sum_j \bm{\sigma}_j \tau^\pm_j ]^{(\lambda)}$,
where $\bm{\sigma}$ is the spin,
$\tau^-$ ($\tau^+$) makes a neutron a proton (vice versa)
and $j$ sums over all nucleons.
The DGT operator can have rank $\lambda=0, 2$, due to symmetry.
For $^{48}$Ca $J^\pi_i = 0^+$ and the DGT strength populates
$J^\pi_f = 0^+,2^+$ states of the final ($f$) $^{48}$Ti.

We perform large-scale shell-model calculations
to study the DGT strength distribution of $^{48}$Ca
with the Lanczos strength function method~\cite{whitehead,caurier-review}. 
This technique is based on acting with the DGT operator
on the initial state of the transition, the $^{48}$Ca ground state.
After typically 300 iterations, 
the states obtained are still not exact eigenstates
of the nuclear interaction but the strength distribution
is a very good approximation to the exact one~\cite{caurier-review}.
When necessary we project into good angular-momentum
diagonalizing the $J^2$ operator. 
The DGT calculations use the
$M$-scheme shell-model code KSHELL \cite{kshell}.
We smear out the final DGT distributions
with Lorentzians of $\Gamma=1$ MeV width
to simulate the experimental energy resolution.

Figure \ref{fig:dgt-interactions} (a) and (b) shows
the DGT strength 
distribution obtained with two different nuclear interactions~\cite{gxpf1b,kb3g}
in the configuration space comprised by one harmonic oscillator major shell
($pf$ shell).
The results in Fig.~\ref{fig:dgt-interactions} (c) use an interaction acting
in two major shells~\cite{iwata-prl2016} ($sd$ and $pf$ shells),
limited to $2\hbar\omega$ excitations.
The three DGT distributions are in reasonable agreement,
suggesting that the theoretical uncertainties due to the
nuclear interaction and the size of the configuration space
are relatively under control.
For instance, the DGT GR centroid energy
only differs by 1.6~MeV
between the two one-major-shell calculations.


{\it DGT GR, pairing and $0\nu\beta\beta$ decay.}
\label{sec:nme}
Next we analyze the properties of the DGT strength distribution.
We probe its dependence on pairing correlations
by adding to the nuclear interaction
\begin{align}
\label{eq:H_t1p}
H' & =  H + G^{JT} P^{JT},
\end{align}
where $P^{J=0,T=1}$ and $P^{J=1,T=0}$
denote the isovector and isoscalar
pairing interactions~\cite{pnpair-poves}, respectively,
with corresponding $G^{01}$ and $G^{10}$ couplings.
For $H$ we take the $pf$-shell GXPF1B interaction, but alternatively using KB3G gives similar results.

\begin{figure}[t]
	\centering
	\includegraphics[width=0.47\textwidth]{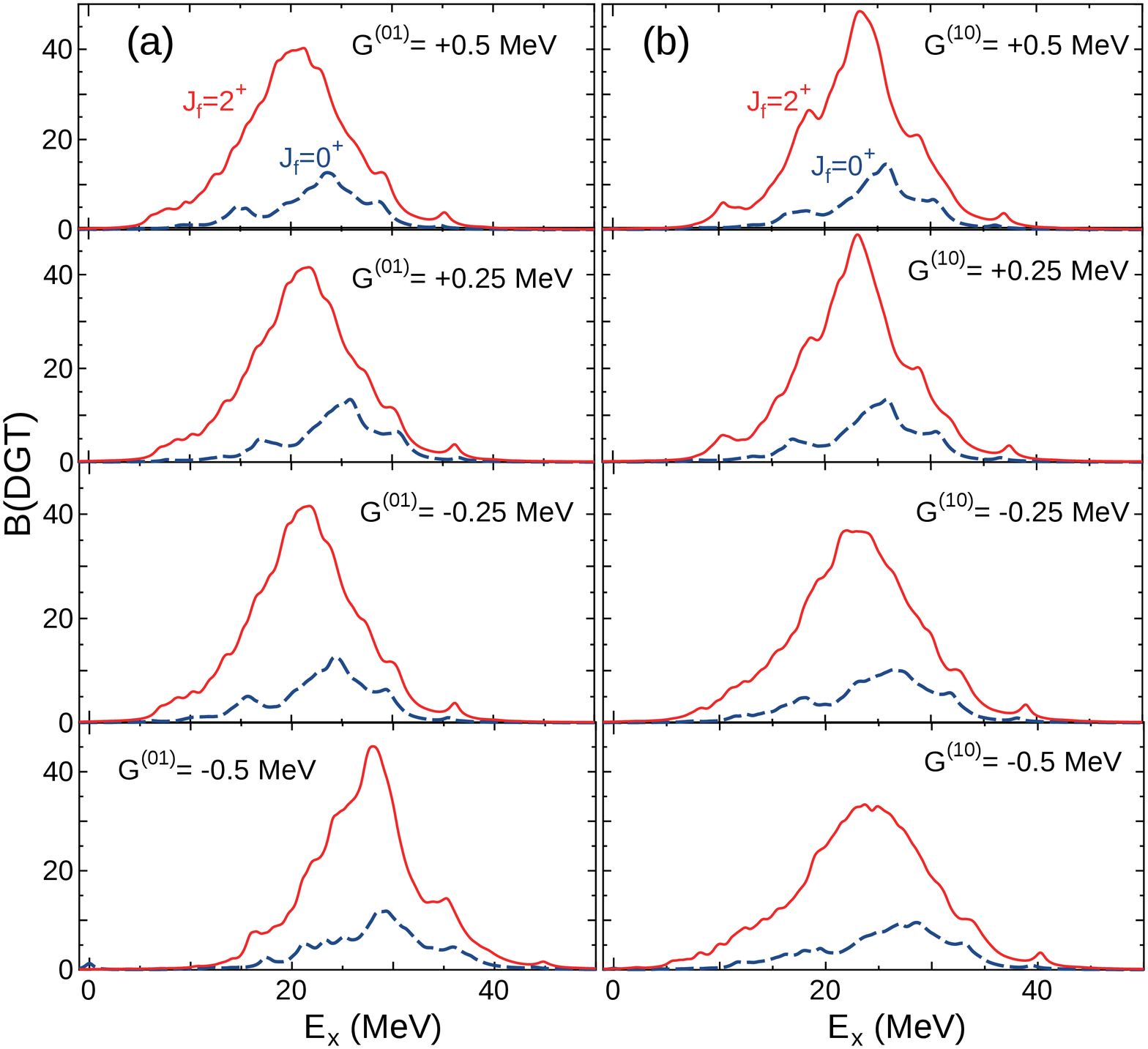}		
	\caption{DGT strength distribution for a range of values
		from 0.5~MeV (top panels) to -0.5~MeV
		(bottom)
		of the couplings added to the GXPF1B interaction:
		(a) $G^{01}$ of the isovector pairing term, 
		(b) $G^{10}$ of the isoscalar pairing term.
		Solid red (dashed blue) lines show $\lambda=2$ ($\lambda=0$) 
		DGT distributions.}
	\label{fig:dgt-t1pair}
\end{figure}

Figure \ref{fig:dgt-t1pair} (a) shows the DGT strength distribution 
for various values, attractive and repulsive,
of the additional isovector pairing term.
The top panel shows the $G^{01}=0.5$~MeV case,
where most of the isovector pairing of the original interaction
is canceled~\cite{dufour-zuker},
while the bottom panel uses
$G^{01}=-0.5$~MeV,
greatly enhancing isovector pairing correlations.
The centroid energy of the DGT distribution,
both for $\lambda=0$ and $\lambda=2$ couplings,
increases with the strength of the isovector pairing interaction, while the DGT GR width remains rather stable.
Likewise Fig.~\ref{fig:dgt-t1pair} (b) shows the DGT strength distribution 
for isoscalar pairing couplings $G^{10}$ ranging from repulsive values
that almost cancel this interaction~\cite{dufour-zuker}
to attractive ones. 
The centroid energy of the DGT GR
is rather independent of the isoscalar pairing coupling.
In contrast, strongly attractive isoscalar pairing 
makes the DGT GR width
broader than with the original nuclear
interaction.

The $0\nu\beta\beta$ decay NMEs 
are also very sensitive to pairing correlations,
both isovector~\cite{caurier-08,barea-09,brown-15}
and isoscalar~\cite{vogel-ip,engel-ip,hinohara-14,menendez-hinohara-2016}.
The $0\nu\beta\beta$ decay NME is given by
a combination of GT, Fermi ($F$) and tensor ($T$) components~\cite{engel-ropp}: 
\begin{align}
    \label{eq:nme}
    M^{0\nu} &= M^{0\nu}_{GT} - \left( \frac{g_V}{g_A} \right)^2 M^{0\nu}_F 
    + M^{0\nu}_T , 
    \\
    M^{0\nu}_{X} &= \langle f| \sum_{jk} \tau^-_j \tau^-_k
    S_{X} 
    V_{X}(r_{jk}) | i \rangle,
\end{align}
where  $g_A/g_V=1.27$ is the ratio of the axial and vector couplings,
the different spin structures are
$S_F=1$, $S_{GT}=\bm{\sigma}_j  \bm{\sigma}_k$ and the tensor $S_{T}$,
and $V_{GT}$, $V_{F}$ and $V_{T}$
are the corresponding neutrino potentials,
which depend on the distance between the decaying neutrons $r_{jk}$.
Equation~(\ref{eq:nme}) uses the closure approximation,
which is accurate to more than 90\%~\cite{horoi-nonclosure-48ca}.
In this approximation
the neutrino potential is the only difference between
the dominant term $M_{GT}^{0\nu}$ and the DGT operator.

\begin{figure}[t]
	\centering
	\includegraphics[width=0.48\textwidth]{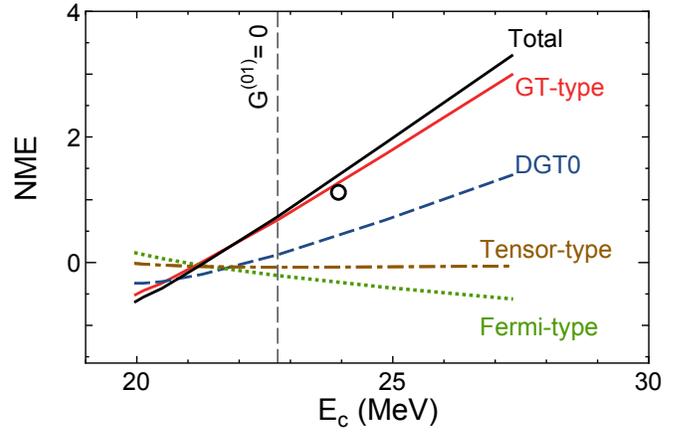}
	\caption{
		$0\nu\beta\beta$ decay NME and DGT GR ($\lambda=2$) centroid energy ($E_c$),
		for the interactions defined in Eq.~(\ref{eq:H_t1p}). 
		The black solid, red solid, green dotted and brown dashed-dotted 
		lines show the total, GT, Fermi and tensor NME parts,
		respectively.
		The blue dashed line denotes
		the DGT transition to the $^{48}$Ti ground state.
		The vertical dashed line
		indicates the results of the original
		GXPF1B interaction.
		The open circle corresponds to the two-shell total result in the $sd$-$pf$ space.
	}
	\label{fig:NME-dgt-Eav}
\end{figure}

We combine the sensitivity to pairing correlations
of the DGT strength distribution, and the well-known sentivity of the
$0\nu\beta\beta$ decay NME to these correlations,	
by studying both observables with $H'$ obtained with various coupling values.
Figure \ref{fig:NME-dgt-Eav} shows NMEs
as a function of the centroid energy of the DGT distribution, defined as
\begin{align}
\label{eq:eav}
E_{\textrm{c}} &= \frac{\sum_f E_f {B(DGT^{-}},i\rightarrow f)}
{\sum_f B({ DGT^{-}},i\rightarrow f)}.
\end{align}
Figure \ref{fig:NME-dgt-Eav} 
highlights that the NME, dominated by $M^{0\nu}_{\rm GT}$,
is well correlated with the average energy of the DGT GR.
This correlation,
driven by the dependence of both observables
to the isovector pairing strength,
agrees well with the results obtained in two major shells,
indicated by an open circle in Fig.~\ref{fig:NME-dgt-Eav}.
This consistency supports the use of the modified interaction $H'$,
which may capture sufficiently well the aspects
relevant for the DGT GR -- $0\nu\beta\beta$ decay
correlation
without the need to reproduce all other nuclear structure properties.
Our study indicates that a measurement of the DGT GR,
besides testing the theoretical calculation,
can provide a hint of the NME value.
A measured centroid energy above (below)
the result of the original nuclear interaction
would suggest a larger (smaller) NME than the initial GXPF1B prediction.
Figure \ref{fig:NME-dgt-Eav}
indicates
that an experimental uncertainty on the DGT GR peak
of a couple of MeV, which might be experimentally accessible 
in the near future~\cite{takaki-cns,uesaka-nppac,cappuzzello},
would be sufficient to shed light
on the $^{48}$Ca $0\nu\beta\beta$ decay NME.
On the other hand, while we find that a narrower DGT GR
is associated with a larger NME,
very small uncertainties below the MeV scale
would be needed to extract information relevant for $0\nu\beta\beta$ decay.

The correlation in Fig.~\ref{fig:NME-dgt-Eav} is
useful if the shell model can reproduce the DGT GR.
This will be tested once DGT data is available.
For the moment, shell model predictions
for the $\rm GT$ strength distribution
of $^{48}$Ca~\cite{caurier-95,iwata-jps}, including the GT GR,
agree quite well with experiment.

{\it DGT and $0\nu\beta\beta$ decay NME.}
\begin{figure}[t]
	\centering
	\includegraphics[angle=0,width=0.48\textwidth]{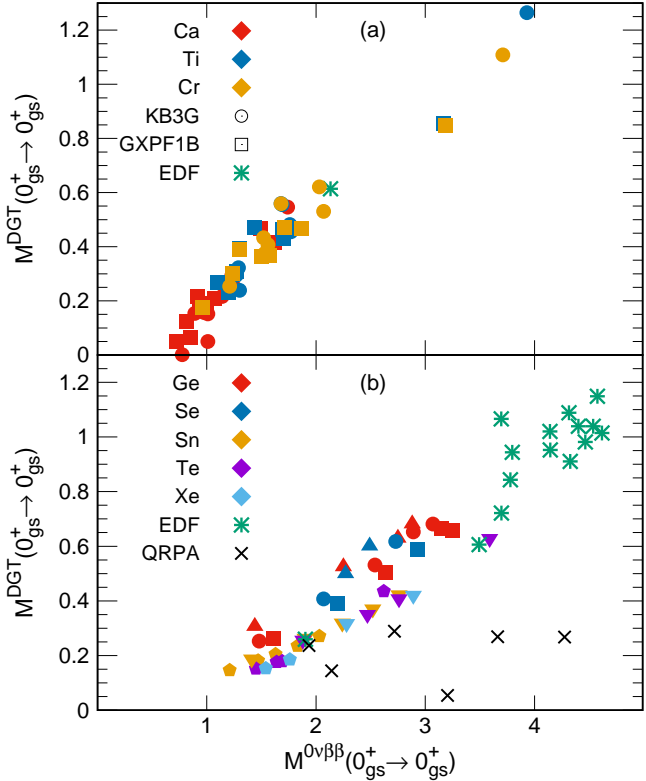}
	\caption{
		Correlation between
		the $0\nu\beta\beta$ decay NME $M^{0\nu}$
		and the DGT matrix element $M^{\rm DGT}$.
		(a) Calcium (red), titanium (blue)
		and chromium (yellow) isotopes
		calculated with the shell model
		GXPF1B (squares) and KB3G (circles) interactions,
		compared to the EDF $^{48}$Ca result~\cite{rodriguez} (green star).
		(b) Germanium (red), selenium (blue), tin (orange),
		tellurium (purple) and xenon (light blue) shell model results
		(filled symbols)
		calculated with the interactions in Refs.~\cite{menendez-09,JUN45,jj44b,qi-12},
		each one represented by a different symbol.
		Compared are EDF~\cite{rodriguez} (green stars) and QRPA~\cite{simkovic-11} (black crosses) results for $\beta\beta$ emitters, and cadmium EDF values.
	}
	\label{fig:NME-linear}
\end{figure}
Figure \ref{fig:dgt-interactions} shows that the
DGT transition into the ground state (gs) of
$^{48}$Ti is a tiny fraction of the total DGT distribution.
Nonetheless this matrix element is expected to be the closest
to $0\nu\beta\beta$ decay since both processes share initial and final states.
We define the DGT matrix element as
\begin{align}
\label{eq:mdgt}
M^{\rm DGT}&=\sqrt{B(DGT^{-}; 0;
0^+_{\rm gs,i}
\rightarrow 
0^+_{\rm gs,f})
} 
\nonumber \\ & 
=
\bigg|
\langle  0^+_{\rm gs,f}\big|\big|
{ \sum_{j,k} [\bm{\sigma}_j \tau^-_j 
	\times \bm{\sigma}_k \tau^-_k ]^{0}}
\big|\big| 0^+_{\rm gs,i}\rangle
\bigg|.
\end{align}
The DGT matrix element is proportional to the
two-neutrino $\beta\beta$ decay matrix element evaluated
in the closure approximation.
The $^{48}\textrm{Ca}$ $M^{\rm DGT}$ shown in Fig.~\ref{fig:NME-dgt-Eav}
is indeed correlated to the $0\nu\beta\beta$ decay NME.

\begin{figure}[t]
	\centering
	\includegraphics[width=0.48\textwidth]{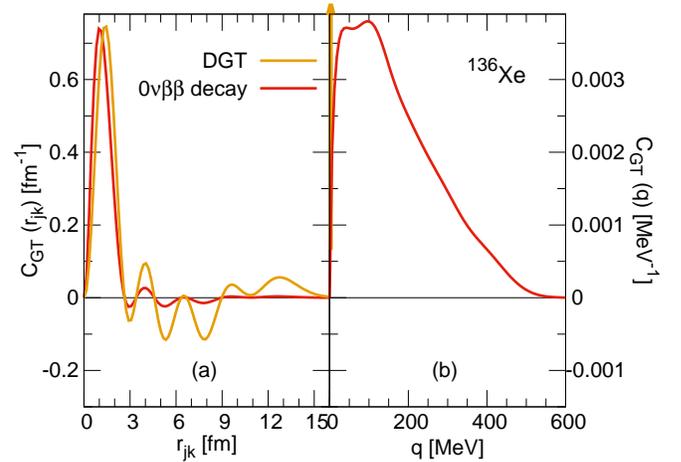}
	\caption{Normalized distributions $C_{GT}$ of the
		$M^{0\nu}_{\rm GT}$ (red) and
		$M^{\rm DGT}$ (orange) matrix elements of $^{136}\textrm{Xe}$
		as a function of (a) the internucleon distance $r_{jk}$,
		(b) the momentum transfer $q$.
		Results obtained with the nuclear interaction from Ref.~\cite{menendez-09}.
	}
	\label{fig:NME-density}
\end{figure}
Figure \ref{fig:NME-linear} explores the relation
between $M^{0\nu}$ and $M^{\rm DGT}$ matrix elements
for twenty-six pairs of initial and final nuclei
comprising initial calcium, titanium and chromium isotopes
with mass numbers $42\le A \le 60$ (panel a);
and seventeen initial germanium, selenium, tin, tellurium and xenon isotopes
with masses $76\le A \le 136$ that include five $\beta\beta$ emitters
(panel b).
We have used various one-major-shell nuclear interactions~\cite{menendez-09,JUN45,jj44b,qi-12} in each mass region.
These interactions have been tested in nuclear spectroscopic studies
and reproduce experimental two-neutrino $\beta\beta$ decay matrix elements
and GT strengths to low-lying states
with a renormalization of the $\bm{\sigma}\tau$ operator~\cite{poves-12,horoi_se,horoi_te,horoi_ge}.
Figure~\ref{fig:NME-linear} shows a simple linear relation
between the DGT and $0\nu\beta\beta$ decay matrix elements,
valid up to $M^{0\nu}\simeq5$.
When taking nuclear states truncated in the seniority basis 
(using the code NATHAN~\cite{caurier-review})
the same linear relation extends to $M^{0\nu}\simeq10$.
The correlation is also common to
calculations in one or two major shells
for results in Fig.~\ref{fig:NME-linear} (a).
 
Furthermore, Fig.~\ref{fig:NME-linear} compares the shell model results with
the nonrelativistic energy-density functional (EDF) ones
for $\beta\beta$ decay emitters and cadmium isotopes from Ref.~\cite{rodriguez}.
The two many-body approaches follow a quite similar correlation.
This is very encouraging given the marked differences
between the shell model and EDF
$M^{0\nu}$ values~\cite{menendez-14}.
On the contrary, the quasiparticle random-phase approximation (QRPA)
calculations for $\beta\beta$ decay emitters from Ref.~\cite{simkovic-11}
give small $M^{DGT}\lesssim 0.4$ matrix elements independently of the
associated $0\nu\beta\beta$ decay NME values.

In order to understand the connection between the two processes,
Fig.~\ref{fig:NME-density} (a) shows the matrix element distributions
as a function of the distance between the transferred or decaying nucleons~\cite{simkovic-08}.
$^{136}$Xe is chosen as an example.
Both matrix elements are dominated by short internucleon distances. In the case of DGT transitions this is because
the intermediate- and long-range contributions cancel
to a good extent.
Radial distributions in the other DGT matrix elements we have studied 
can be somewhat different,
but the approximate cancellation between intermediate and long internucleon distances
is systematically observed.
By contrast, Fig.~\ref{fig:NME-density} (b) shows that
the momentum transfers are quite different,
vanishing for DGT transitions and
peaking around $100$~MeV in $0\nu\beta\beta$ decay. 

The short range character of both DGT and $0\nu\beta\beta$ decay matrix elements
can explain the simple linear relation between them.
References~\cite{bogner-10,bogner-12} showed that
if an operator only probes the short-range physics of low-energy states,
the corresponding matrix elements factorize
into a universal operator-dependent constant
times a state-dependent number common to all short-range operators.
A linear relation
between the DGT and $0\nu\beta\beta$ decay matrix elements follows.
Our correlation depends moderately on the mass region probably
because of the approximate cancellation
of intermediate- and long-range contributions in the DGT matrix elements.
This explanation is consistent with the different pattern of the QRPA results,
as QRPA DGT transitions
do not show any cancellation between intermediate and long internucleon distances~\cite{simkovic-11}, contrary to the shell model.

Another difference
between shell model and QRPA DGT matrix elements
appears when Eq.~(\ref{eq:mdgt}) is evaluated
introducing a complete set of intermediate states.
While in the QRPA intermediate $1^+$ states up to 15~MeV
can be relevant~\cite{simkovic-11},
typically canceling low-energy contributions,
in the shell model the impact of $1^+$ states beyond 8~MeV is minor~\cite{poves-12}.
A possible reason is that shell model calculations, except for $^{48}$Ca, miss spin-orbit partners needed to reproduce the GT GR
at 10-15~MeV.
This difference may be connected to the disagreement between QRPA and shell model
results in Fig.~\ref{fig:NME-linear}.
On the other hand,
the good correlation between shell model
DGT and $0\nu\beta\beta$ decay matrix elements
suggests that intermediate states may not be decisive,
because DGT transitions involve $1^+$ states only,
while $0\nu\beta\beta$ decay is dominated by intermediate states
with other angular momenta~\cite{horoi_se,horoi_te,horoi_ge,simkovic-08}.

The linear correlation between the DGT transition
and $0\nu\beta\beta$ decay in Fig.~\ref{fig:NME-linear} could be used
to constrain the value of NMEs from DGT experiments.
However measuring the $M^{\rm DGT}$
to the ground state is a formidable challenge.
Our calculations predict that this transition
accounts for only about 0.03 per mil of the total DGT strength.
Clean experiments as well as elaborate analysis
are necessary~\cite{cappuzzello}.

Corrections to the correlations found in this work
may arise from the renormalization
of the $0\nu\beta\beta$ or DGT operators
due to missing many-body correlations~\cite{bertsch-82,arima-87,towner-97},
two-body currents~\cite{menendez-11},
or an axial tensor polarizability~\cite{Shanahan17}. 
Nonetheless updated correlations could be obtained
adding the new contributions
to the matrix elements defined in Eqs.~(\ref{eq:nme}) and (\ref{eq:mdgt}).

{\it Summary.}
\label{sec:summary}
We have performed large-scale shell model calculations
that suggest that double charge-exchange reactions
can be a very valuable tool to constrain $0\nu\beta\beta$ decay NMEs.
Our study indicates that
if theoretical calculations reproduce upcoming DGT transition data
as they do with GT strength distributions,
the experimental energy of the DGT GR
will inform on the $^{48}$Ca $0\nu\beta\beta$ decay NME value.
More generally, we have found a linear correlation between
the DGT transition to the final nucleus ground state 
and the $0\nu\beta\beta$ decay NME.
The correlation originates in the dominant short-range character
of both transitions.
Recent and planned experiments looking for DGT strength distributions 
will test our theoretical predictions
and open the door to constraining $0\nu\beta\beta$ decay NMEs
in double charge-exchange experiments.

{\it Acknowledgments.}
We thank Scott K. Bogner,  Yutaka Utsuno, and Motonobu Takaki
for very stimulating discussions.
This work was supported by JSPS KAKENHI Grant (25870168, 17K05433) and 
CNS-RIKEN joint project for large-scale nuclear structure calculations.
It was also supported by MEXT and JICFuS as a priority issue 
(Elucidation of the fundamental laws and evolution of the universe, hp170230) 
to be tackled by using Post K Computer.


\end{document}